\global\def\draftcontrol{0}

   \def\versionno{ holomorphic anomaly }

\catcode`\@=11

\expandafter\ifx\csname draftcontrol\endcsname\relax\global\def\draftcontrol{0} 
\fi 

{\count255=\time\divide\count255 by 60 
\xdef\hourmin{\number\count255} 
\multiply\count255 by-60\advance\count255 by\time 
\xdef\hourmin{\hourmin:\ifnum\count255<10 0\fi\the\count255}} 
\def\draftdate{\number\month/\number\day/\number\year\ \ \ \hourmin } 

\newcommand\makepapertitle{\par

  \begingroup 
    \renewcommand\thefootnote{\@fnsymbol\c@footnote}%
    \def\@makefnmark{\rlap{\@textsuperscript{\normalfont\@thefnmark}}}%
    \long\def\@makefntext##1{\parindent 1em\noindent 
            \hb@xt@1.8em{%
                \hss\@textsuperscript{\normalfont\@thefnmark}}##1}%
     \newpage 
     \global\@topnum\z@   
     \@makepapertitle 
     \thispagestyle{empty}\@thanks 
  \endgroup 
  \setcounter{footnote}{0}%
  \global\let\thanks\relax 
  \global\let\makepapertitle\relax 
  \global\let\@makepapertitle\relax 
  \global\let\@thanks\@empty 
  \global\let\@author\@empty 
  \global\let\@date\@empty 
  \global\let\@title\@empty 
  \global\let\title\relax 
  \global\let\author\relax 
  \global\let\date\relax 
  \global\let\and\relax 
  \def\version{\let\version\@version\@gobble} 
} 
\def\@makepapertitle{%
  \newpage 
   \ifnum\draftcontrol=1 {} 
   \version\versionno 
   \vskip 7em%
   \else 
   \hfill\hbox to 3cm {\parbox{4cm}{\@pubnum}\hss}%
   \vskip 7em%
   \fi 
   \begin{center}%
   \let \footnote \thanks 
      {\hskip -0\textwidth \hbox to 1\textwidth%
        {\centerline{\Large\bf{\noindent\@title}}}}%
     \vskip 1.5em%
     {\normalsize
       \lineskip 1.5em%
       \begin{tabular}[t]{c}%
         \@author 
       \end{tabular}\par}%
     \vskip 5.5em%
     {\@bstract}%
     \end{center}%
     \vfill
     \@date%
     \vskip 1.5em%
   \par 
} 

\gdef\@pubnum{} 
\def\pubnum#1{%
  \gdef\@pubnum{#1}} 

\gdef\@bstract{} 
\def\Abstract#1{%
  \gdef\@bstract{%
   \parbox{\textwidth-5pc}{%
   \centerline{\bf Abstract}\penalty1000 
   \renewcommand\baselinestretch{1.0} 
   \noindent
   {#1}}} 
} 

\gdef\@email{}
\def\email#1{%
   \gdef\@email{%
   Email: {\tt #1}}
}

\def\ps@paper{\let\@mkboth\@gobbletwo%
     \ifnum\draftcontrol=1 
        \def\@oddfoot{\hbox to \textwidth{\tiny \versionno \hfil\tiny\draftdate}%
        \hskip -\textwidth \hbox to \textwidth{\hfil\rm\thepage\hfil}}%
     \else\def\@oddfoot{\hbox to \textwidth{\hfil\rm\thepage\hfil}} 
     \fi 
     \let\@evenfoot\@oddfoot 
} 

\def\body{\clearpage 
          \pagestyle{paper} 
        } 
\newenvironment{acknowledgments}{%
\vskip 3.25ex 
\noindent {\bf Acknowledgments} 
} 

\def\@version#1{\ifnum\draftcontrol=1 
\typeout{}\typeout{#1}\typeout{} 
\vskip3mm\centerline{\hbox{\fbox{\normalsize{\tt DRAFT -- #1 -- } 
                   {\draftdate}}}}\vskip3mm 
\fi} 
\let\version\@version 
\long\def\eqlabel#1{\ifnum\draftcontrol=1 
                    \tag@false  
                    \tag*{(\theequation) \hbox to -0.2cm{\hspace{0cm}\small{#1}\hss}} 
                    \refstepcounter{equation}  
                    \edef\@currentlabel{\theequation} 
                    \ltx@label{#1}          
                    \else 
                    \label{#1} 
                    \fi 
                    } 
\let\st@bibitem\@bibitem 
\let\st@lbibitem\@lbibitem 
\ifnum\draftcontrol=1 
  \def\@bibitem#1{%
    \st@bibitem{#1}\a@@label{#1}\ignorespaces} 
  \def\@lbibitem[#1]#2{%
    \st@lbibitem[#1]{#2}\a@@label{#2}\ignorespaces} 
  \def\a@@label#1{%
    \gdef\a@lab{\smash{\normalfont\small#1}} 
    \ifvmode 
      \if@inlabel 
        \global\setbox\@labels\hbox{%
          \llap{\a@lab\let\a@lab\relax 
                \kern\@totalleftmargin\kern\marginparsep}%
          \box\@labels}%
      \fi 
    \fi} 
\fi 

\documentclass[letterpaper,12pt]{article} 

\usepackage{amsmath,bm,amsfonts,amssymb,array,calc,amsthm,rotating}
\usepackage{color}
\usepackage[colorlinks=false]{hyperref}

\renewcommand\baselinestretch{1.25} 
\definecolor{refcol}{rgb}{0.2,0.2,0.8}
\definecolor{eqcol}{rgb}{.6,0,0}
\definecolor{purple}{cmyk}{0,1,0,0}

\gdef\@citecolor{refcol}
\gdef\@linkcolor{eqcol}
\def\colorlinkspurple{\gdef\@urlcolor{purple}}
\def\colorlinksblue{\gdef\@urlcolor{blue}}
\def\colorlinksred{\gdef\@urlcolor{red}}

\catcode`\@=12 

\begin{document} 


\def\F{\mathcal F}


\title{
\parbox{\textwidth}{\begin{center}Comments on the Holomorphic Anomaly\\
in Open Topological String Theory\end{center}}}

\pubnum{%
CALT-68-2651}
\date{June 2007}

\author{
Paul L. H. Cook, Hirosi Ooguri and Jie Yang \\[0.5cm]
\it California Institute of Technology, Pasadena, CA 91125, USA
}

\Abstract{We show that a general solution to the extended holomorphic anomaly equations
for the open topological string on D-branes in a Calabi-Yau manifold, recently written
down by Walcher in arXiv:0705.4098, is obtained from the general solution to the holomorphic anomaly equations for the 
closed topological string on the same manifold, by shifting the closed string 
moduli by amounts proportional to the 't Hooft coupling.}

\makepapertitle

\body

\version\versionno

\vskip 1em

\newpage

Recently Walcher \cite{WalcherTP} showed that topological string amplitudes with D-branes in compact Calabi-Yau manifolds satisfy a set of differential equations, which generalize the holomorphic anomaly equations of \cite{BCOV} (henceforth referred to as BCOV). We would like to point out that a general solution to Walcher's equations is simply related to a general solution to the original holomorphic anomaly equation for closed topological string theory on the same manifold.

Walcher's extended holomorphic anomaly equation is,
\begin{eqnarray}
\bar \partial_{\bar i}{\mathcal F}^{(g, h)}_{i_1, \cdots, i_n}&=&\frac{1}{2}\sum_{\begin{subarray}{l} g_1+g_2=g\\ h_1+h_2=h\end{subarray}}
\overline{C}^{jk}_{\bar i} \sum\limits_{s, \sigma}\frac{1}{s!(n-s)!}{\mathcal F}^{(g_1, h_1)}_{ji_{\sigma(1)}, \cdots, i_{\sigma(s)}}
{\mathcal F}^{(g_2, h_2)}_{ki_{\sigma(s+1)}, \cdots, i_{\sigma(n)}}
+\frac{1}{2}\overline{C}^{jk}_{\bar i}{\mathcal F}^{(g-1, h)}_{jki_1, \cdots, i_n}\nonumber\\
&&-\Delta^j_{\bar i}{\mathcal F}^{(g, h-1)}_{ji_1, \cdots, i_n}
-(2g-2+h+n-1)\sum^n_{s=1}G_{i_s\bar i}{\mathcal F}^{(g, h)}_{i_1 ,\cdots, i_{s-1}, i_{s+1}, \cdots, i_n},\label{Walcher}
\end{eqnarray}
derived under the assumption that the topological string amplitudes do not depend on any continuous open string moduli \cite{WalcherTP}. This equation is valid for $(2g-2+h+n) > 0$, except for ${\mathcal F}^{(1,0)}_i$ and ${\mathcal F}^{(0,2)}_i$ for which there are additional terms in the equation, which we will also take into account below. The ingredients are $\F^{(g,h)}_{i_1,...,i_n}$, which are topological string amplitudes with worldsheet genus $g$, $h$ boundaries and $n$ insertions of closed string marginal operators indexed by $i_1, \cdots ,i_n$; $\overline{C}_{\bar i}^{jk} = \overline{C}_{\bar i \bar j \bar k}e^{2K} G^{j \bar j} G^{k \bar k}$, where $\overline{C}_{\bar i \bar j \bar k}$ is the Yukawa coupling and indices are raised and lowered using the Zamolodchikov metric $G_{i \bar j}=\partial_i \partial_{\bar j} K$; and $\Delta_{\bar i}^j =e^K G^{j \bar k} \Delta_{\bar i \bar k}$, where $\Delta_{\bar i \bar k}$ is the disk amplitude with two insertions. Note that these are different to the $\Delta$ (with or without indices) that appear in BCOV, which we will denote as $\hat\Delta$ below. This equation provides a recursion relation for open topological string amplitudes in terms of contributions of lower genus $g$ or boundary number $h$. Since this equation specifies only the anti-holomorphic dependence of each amplitude, there is an additional holomorphic ambiguity, consisting of a holomorphic function at each genus and each number of boundaries.

Following BCOV, we define the generating function for open topological string amplitudes,
\begin{eqnarray}
\nonumber
W(x, \varphi; t, \bar t)= \sum_{g,h,n}\frac{1}{n!}\lambda^{2g-2}\mu^h {\mathcal F}^{(g, h)}_{i_1, \cdots, i_n} x^{i_1}\cdots x^{i_n}\left(\frac{1}{1-\varphi}\right)^{2g-2+h+n} && \\
+\left(\frac{\chi}{24} - 1-\frac{N}{2} \lambda^{-2} \mu^2 \right) \log \left(\frac1{1-\varphi}\right),\label{generating}&&
\end{eqnarray}
where the sum is over $g, h, n \geq 0$ such that $(2g-2+h+n) > 0$, $\lambda$ is the topological string coupling constant, and $\mu$ is the 't Hooft coupling constant, namely $\lambda$ times the topological string Chan-Paton factor. In the last term on the right, $\chi$ is the Euler characteristic of the Calabi-Yau manifold and $N$ is the number of open string ground states with zero charge. This term contributes to the holomorphic anomaly equations for ${\mathcal F}_i^{(1,0)}$ and ${\mathcal F}_i^{(0,2)}$, reproducing (3.10) of BCOV and (2.87) of Walcher respectively. The generating function $W$ satisfies an extension of BCOV's equation (6.11) by a $\mu$-dependent term, namely,
\begin{equation}
\label{open_generatingequation}
\frac{\partial}{\partial \bar t^{\bar i}}e^{ W(x, \varphi; t, \bar t)}=\left(\frac{\lambda^2}{2}\overline{C}^{jk}_{\bar i}\frac{\partial^2}{\partial x^j\partial x^k}
-G_{\bar i j}x^j\frac{\partial}{\partial\varphi}-\mu \Delta^j_{\bar i}\frac{\partial}{\partial x^j}\right)e^{ W(x, \varphi; t, \bar t)},
\end{equation}
which reproduces the open topological string holomorphic anomaly equation (\ref{Walcher}) for each genus and boundary number.

Our key result is that equation (\ref{open_generatingequation}) can be rewritten in the same form as the closed topological string analogue by simply shifting
\begin{equation}
\label{shift}
x^i \rightarrow x^i + \mu\Delta^i, \;\;\; \varphi \rightarrow \varphi + \mu\Delta,
\end{equation}
where $\Delta^i$ and $\Delta$ are defined modulo holomorphic ambiguities by $ \Delta_{\bar i \bar j} = e^{-K} G_{\bar j k} \partial_{\bar i} \Delta^k
     = e^{-K} D_{\bar i} D_{\bar j} \Delta$. After this shift equation (\ref{open_generatingequation}) becomes,
\begin{equation}
\label{shifted_generatingequation}
\frac{\partial}{\partial \bar t^{\bar i}}e^{W(x+\mu\Delta, \varphi+\mu\Delta; t, \bar t)}=\left(\frac{\lambda^2}{2}\overline{C}^{jk}_{\bar i}\frac{\partial^2}{\partial x^j\partial x^k}
-G_{\bar i j}x^j\frac{\partial}{\partial\varphi} \right)e^{W(x+\mu\Delta, \varphi+\mu\Delta; t, \bar t)}.
\end{equation}
This is exactly the same as BCOV's original equation (6.11) for the closed topological string, with the $\mu$-dependent term absorbed by means of the shift (\ref{shift}).

Our result follows from a straightforward application of the chain rule: noting that $\bar\partial_{\bar i} \Delta^j = \Delta_{\bar i}^j$, the variable shift produces two new terms on the left,
\begin{displaymath}
  \left(\mu\Delta_{\bar i}^j \frac\partial{\partial x^j} + \mu \Delta_{\bar i} \frac\partial{\partial\varphi} \right) e^{W}.
\end{displaymath}
The first is the additional $\mu$-dependent term on the right of (\ref{open_generatingequation}). Using $G_{\bar i j}\Delta^j = \Delta_{\bar i}$, the second term combines with the second term on the right of (\ref{shifted_generatingequation}) to give $-G_{\bar i j}(x^j+\mu\Delta^j)\frac\partial{\partial\varphi} e^{W}$, which is required for matching powers of $x+\mu\Delta$ in the expansion of the generating function. Thus we have reproduced the open topological string holomorphic anomaly equations from the closed topological string holomorphic anomaly equations, simply by a shift of variables.

An immediate consequence of this is a general proof of the Feynman rule method of solving the open topological string anomaly equations appearing in subsection 2.10 of Walcher. Since our shifted $W$ satisfies the closed string differential equation (\ref{shifted_generatingequation}), the proof of the closed string Feynman rules presented in subsection 6.2 of BCOV applies immediately. The shift has, in fact, an elegant interpretation in terms of the Feynman rules. Equation (6.12) in BCOV defines the function,
\begin{equation}
  \label{closed_Y}
  Y(x,\varphi;t,\bar{t}) = -\frac1{2\lambda^2} (\hat\Delta_{ij} x^i x^j + 2 \hat\Delta_{i\varphi} x^i \varphi + \hat\Delta_{\varphi\varphi} \varphi^2) + \frac12 \log\left(\frac{\det \hat\Delta}{\lambda^2} \right),
\end{equation}
where the $\hat\Delta_{ij}$ are the inverses of the corresponding propagators $S^{ij}$. Expanding $Z=\int dxd\varphi \exp(Y+W)$ in powers of $\lambda$ then produces the full Feynman diagram expansion of the closed topological string amplitudes. The shift (\ref{shift}) produces the additional terms appearing in the open string Feynman diagrams, shown in subsection 2.10 of Walcher. In field theory language, the shift effectively generates the vacuum expectation values $\langle x^i \rangle = \Delta^i$ and $\langle \varphi \rangle = \Delta$, and so terms containing $\Delta^i$ and $\Delta$ correspond to diagrams with tadpoles.

The simple reformulation of the open string anomaly in terms of the closed string anomaly should also make it possible to apply Yamaguchi and Yau's \cite{Yamaguchi} reformulation of the closed string amplitude diagram expansion to the open string case, which would give a computationally more tractable formulation than the Feynman diagram rules used here.

This open-closed relationship is reminiscent of large $N$ duality, where the background is shifted by an amount proportional to the 't Hooft coupling. It would be interesting to explore the implications of this for the Gromov-Witten and Gopakumar-Vafa invariants.

\begin{acknowledgments}
This research is supported in part by DOE grant DE-FG03-92-ER40701. 
H.O. thanks the Galileo Galilei Institute for Theoretical Physics for 
hospitality and the INFN for partial support during the completion of this work. 
\end{acknowledgments}

\end{document}